\pdfoutput=1
\documentclass[10pt,amsmath,amssymb,aps,prl,superscriptaddress,twocolumn,floatfix]{revtex4-2}
\usepackage{graphicx} 
\usepackage{wrapfig}
\usepackage{bm}
\usepackage[dvipsnames]{xcolor}
\usepackage[bookmarks=false,linkcolor=NavyBlue,urlcolor=NavyBlue,colorlinks,citecolor=NavyBlue]{hyperref}
\usepackage{subfigure}

\definecolor{darkblue}{HTML}{0b3873}

\newcounter{para}
\newcommand{\para}{\par\refstepcounter{para}\textbf{{\color{cyan}[\thepara]}}\space}

\usepackage[normalem]{ulem}

\newcommand{\xCornell}{Department of Physics, Cornell University, Ithaca, NY 14853, USA}

\newcommand{\xEwha}{Department of Physics, Ewha Womans University, Seoul, South Korea\\\vspace{0.7em}}

\newcommand{\figref}[2]{\hyperref[#1]{\ref{#1}(#2)}}
\newcommand{\figsref}[2]{\hyperref[#1]{\ref{#1}#2}}

\begin{document}
\let\para\relax
\title{Graphlet Histogram Representation Database of Inorganic Crystals}
\author{Aaditya Panigrahi}
\affiliation{\xCornell}
\author{Yanjun Liu}
\affiliation{\xCornell}
\author{Omri Lesser}
\affiliation{\xCornell}
\author{Krishnanand Mallayya}
\affiliation{\xCornell}
\author{Eun-Ah~Kim}
\affiliation{\xCornell}
\affiliation{\xEwha}
\begin{abstract}
Machine learning models for materials property prediction increasingly rely on representations learned end-to-end from large density-functional-theory databases, limiting their applicability when only scarce experimental data are available.
Domain-knowledge-driven representations precomputed from crystal structures alone offer a data-efficient, interpretable alternative, but existing approaches capture at most composition or bonding connectivity and discard local structural geometry.
Here, we present Graphlet-MP, a database of graphlet histogram representations for 149{,}082 inorganic crystals from the Materials Project (MP).
Seventy-nine distributions describe each material over three hierarchical graphlet orders---atomic sites, bonded pairs, and bond-angle triplets---extracted via screened Voronoi tessellation from the crystallographic information file.
We provide a complete technical specification of the representation, an Earth Mover's Distance metric for comparing materials in this space, and the full precomputed database.
An accompanying open-source codebase enables users to generate graphlet histograms for arbitrary crystal structures, including experimentally determined ones, and to extend the database to new materials or target properties.
\end{abstract}
\maketitle

\begin{figure*}[t]
    \centering
    \includegraphics[width=0.7\textwidth]{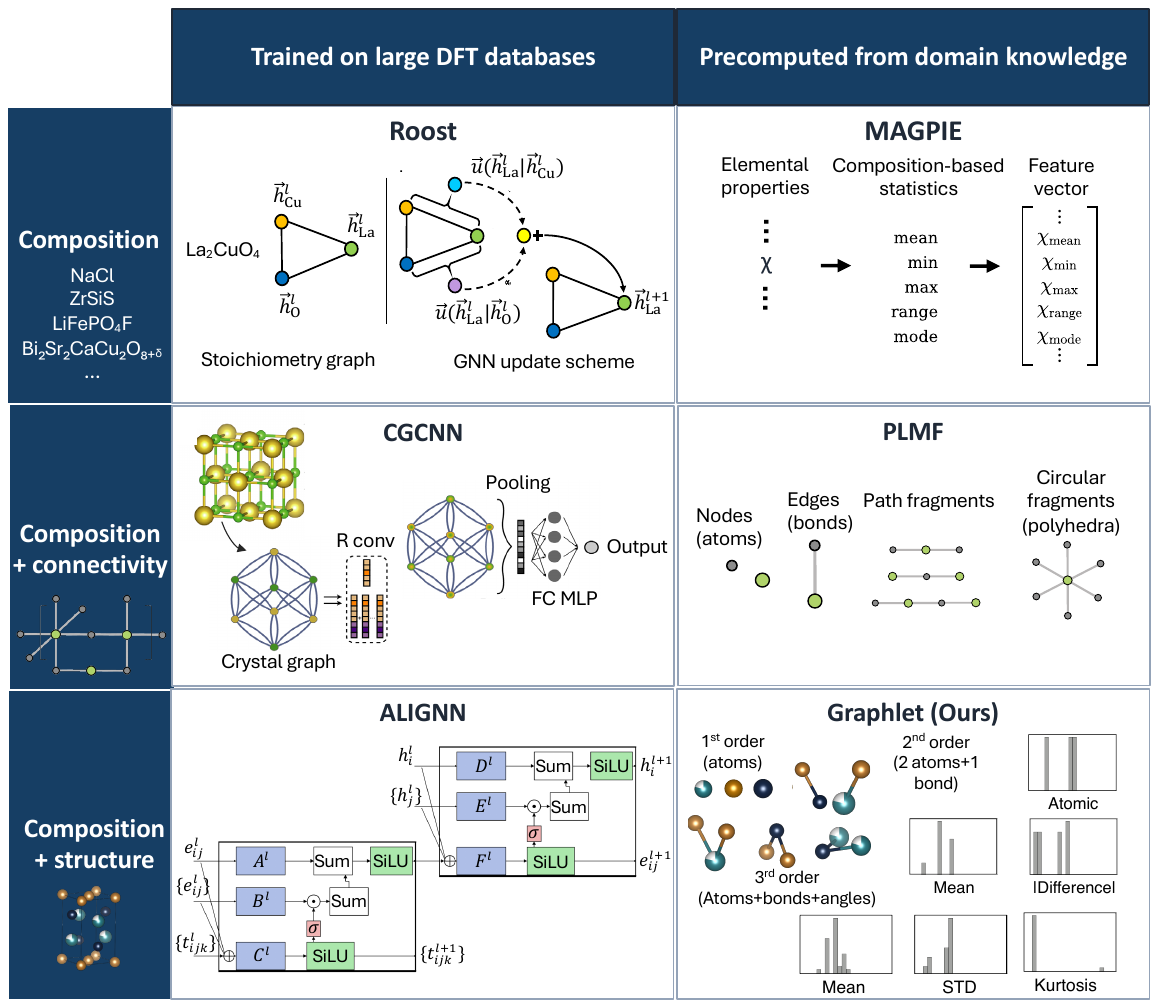}
    \caption{\textbf{Landscape of representations for crystal property prediction.}
    \textbf{Roost}~\cite{goodall_predicting_2020}: elemental embeddings $\vec{h}^l_X$ updated via attention-weighted perturbations $\vec{u}(\vec{h}^l_X|\vec{h}^l_Y)$.
    \textbf{MAGPIE}~\cite{meredig_combinatorial_2014,ward_general-purpose_2016}: composition-weighted statistics (mean, min, max, range, mode) of elemental properties $\chi$.
    \textbf{CGCNN}~\cite{xie_crystal_2018}: $R$ graph-convolutional layers on the crystal graph, followed by pooling.
    \textbf{PLMF}~\cite{isayev_universal_2017}: aggregate statistics over path and circular fragments of the periodic bond graph.
    \textbf{ALIGNN}~\cite{choudhary_atomistic_2021}: interleaved message passing on the bond graph (atom features $h^l_i$, bond features $e^l_{ij}$) and its line graph (bond-angle features $t^l_{ijk}$).
    \textbf{Graphlet} (this work): hierarchical histograms over sites (1st order), bonded pairs (2nd order), and bond-angle triplets (3rd order), yielding 79 distributions per material.}
    \label{fig:intro}
\end{figure*}

\setcounter{dbltopnumber}{2}
\para
Data-driven approaches promise to accelerate the discovery of functional materials, but their success hinges critically on effective material representation.
The prevailing pursuit of learning the representation and property prediction simultaneously end-to-end~\cite{choudhary_atomistic_2021,goodall_predicting_2020,xie_crystal_2018} demands tens to hundreds of thousands of labeled examples.
Nevertheless, the most consequential materials discoveries---high-temperature superconductors, novel piezoelectrics, next-generation dielectrics---are often in regimes where labeled experimental data are precious and hard-won; the largest curated experimental superconductor structure dataset, for example, contains only ${\sim}4{,}350$ superconductors with representative structural information~\cite{sommer_3dsc_2023,lesser_learning_2025} \footnote{There are ${\sim}9{,}150$ entries in 3DSC. Out of those, 6{,}463 are unique superconductors, and only 4{,}325 have unique representative structural information, as measured by the graphlet histogram earth mover distance metric, as of June 2026~\cite{lesser_learning_2025}.}.
Researchers therefore turn to high-throughput density functional theory (DFT) calculations, but DFT deviates from experiment in uncalibratable ways---thereby propagating systematic errors into downstream models. Moreover, most such approaches further omit or approximate structural information (Fig.~\ref{fig:intro}, left column).
Composition-only representations introduced in Refs.~\cite{goodall_predicting_2020,wang_compositionally_2021} cannot distinguish polymorphs.
Graph neural networks (GNNs) either omit bond angles~\cite{xie_crystal_2018} or encode them opaquely~\cite{choudhary_atomistic_2021}, while requiring large DFT datasets to learn the representation.

\para Alternatively, one could anchor representations in domain knowledge, explicitly embedding physical invariances and chemical intuition rather than learning them from data~\cite{ghiringhelli_big_2015}, separating representation from the property predictor training.
This principle was recognized early in materials informatics, but comprehensively capturing structural information alongside composition has remained an open challenge.
MAGPIE~\cite{meredig_combinatorial_2014,ward_general-purpose_2016} takes this route at the composition level, compressing all elemental properties into scalar statistics (mean, min, max, range, mode) over the unit cell; it retains no information about which atoms neighbor which, and therefore cannot distinguish polymorphs.
PLMF~\cite{isayev_universal_2017} extends to the connectivity level by cataloging path fragments and circular fragments of the bond graph. Still, it also reduces the fragment population to aggregate statistics and does not encode bond angles.

\para In Ref.~\cite{lesser_learning_2025}, we introduced and employed the graphlet histogram representation that integrates compositional and structural information, which served as input to a Gaussian-process model (GP-$T_c$) for predicting the superconducting transition temperature.
Without the burden of learning the representation, we trained GP-$T_c$ interpretably to achieve $R^2 = 0.93$ for predicting the superconducting transition temperature, using only ${\sim}4{,}350$ structurally distinct, experimentally obtained training data.

GP-$T_c$'s interpretability enabled an aggressive compression of the 67-distribution feature space to just four descriptors, revealing that the distribution of electron-affinity differences between neighboring atoms is the single most informative predictor of $T_c$ across chemically diverse superconducting families.
GP-$T_c$ further demonstrated experimental actionability: it predicted, and the authors experimentally confirmed, superconductivity in PtPb$_3$Bi ($T_c\approx3$~K).
These results validate graphlet histograms as a powerful representation for materials discovery. However, providing precise documentation of the construction of the graphlet histogram representation and the associated distance metric as a standalone resource would benefit the broader community, given the representation's versatility.

\para
The present work fills this gap.
We provide a complete technical specification of the graphlet histogram representation, accompanied by Graphlet-MP, a precomputed database of 149{,}082 materials.
We compute the representation using structures from the Materials Project (MP), a comprehensive, open-access community resource.
By publishing this database, we aim to make the graphlet histogram representation broadly accessible, enabling researchers to develop predictors for diverse target properties, including superconductivity, dielectricity, and piezoelectricity. As experimental structure databases are behind paywalls, we build Graphlet-MP on DFT-calculated structures in MP. However, we also supply an open-source codebase~\cite{source_code}
for users to generate graphlet histograms from any Crystallographic Information Files (CIFs). With this code, users can convert private, high-value CIF files into graphlet histograms for downstream tasks such as property prediction or structural similarity comparison.
In the rest of this paper, we define the graphlet histogram representation, specify the screened Voronoi neighbor construction, introduce the Earth Mover's Distance metric used to compare graphlet distributions, and describe the resulting precomputed database.

\begin{figure*}[t]
    \centering
    \includegraphics[width=0.7\textwidth]{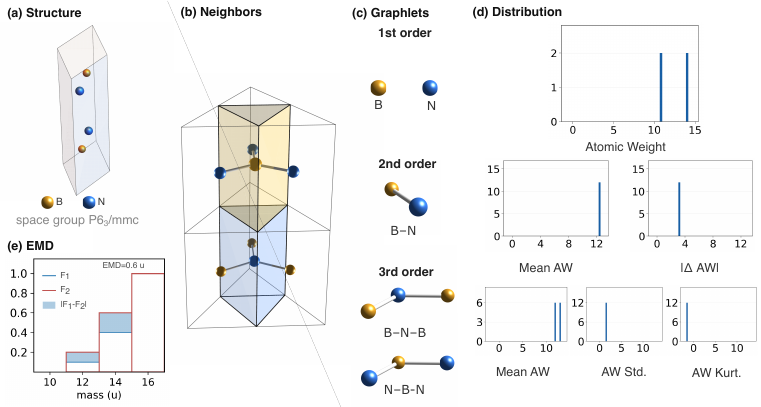}
    \caption{\textbf{Graphlet featurization pipeline illustrated on hexagonal boron nitride (h-BN, mp-984).}
    \textbf{(a)} Crystal structure (space group P6$_3$/mmc) with B (gold) and N (blue) sites in the primitive unit cell.
    \textbf{(b)} Screened Voronoi neighbor graph; lines connect atoms whose Wigner--Seitz cells share a face within the covalent-radius cutoff.
    \textbf{(c)} Graphlet hierarchy: 1st-order (individual sites), 2nd-order (bonded pairs), and 3rd-order (center-site triplets) subgraphs extracted from the neighbor graph.
    \textbf{(d)} Atomic-weight distributions at each graphlet order, illustrating the histogram representation stored in the database for each of the 79 features.
    \textbf{(e)} Earth Mover's Distance between two 1st-order atomic-weight distributions, shown as the area between their cumulative distribution functions.}

    \label{fig:graphlets}
\end{figure*}

\para The Graphlet histogram representation describes a crystalline material through the explicit distribution of elemental and structural properties over localized geometric motifs, or graphlets, within the unit cell.
While graph neural network-based approaches treat the crystal graph as a scaffold for recursive message passing to learn opaque latent embeddings, here we extract deterministic, interpretable subgraphs.
These graphlets systematically capture the local environment in increasing order: first-order graphlets capture the identity and electronic state of individual atomic sites; second-order graphlets capture connectivity and bond-level properties; and third-order graphlets capture angular constraints around each bond.
While our systematic principle for the graphlet construction captures local environments in the crystal comprehensively, it yields wildly different numbers of graphlets across crystals. To treat all materials using a standardized data structure, we bin graphlet properties into histograms.
When a task requires fixed-length inputs, a shared binning scheme aligns all materials into a common, fixed-dimensional feature space without sacrificing the interpretability of the underlying physical features.
We illustrate the full pipeline from crystal structure to neighbor graph to deterministic graphlet distributions in Fig.~\ref{fig:graphlets} for hexagonal boron nitride (h-BN, mp-984).

\para We track 10 elemental attributes for each site (Pauling electronegativity, electron affinity, ionization potential, covalent radius, atomic weight, periodic-table column, and s/p/d/total valence electron counts). We source the elemental attributes from CRC Handbook of Chemistry and Physics~\cite{Rumble2020} based on experimental measurements. For non-stoichiometric materials, we assign site-specific attributes by weighting the elemental features of the element occupying the site by its occupancy. For instance, a site with composition 0.5\,Li + 0.5\,Na would be assigned the attributes 0.5\,Li + 0.5\,Na. We use these site-specific attributes to form the first-order graphlet histograms, calculated as counts of the attribute values across all sites in the crystal. See the first row in Fig.~\ref{fig:graphlets}d for an illustration of the first-order graphlet histogram, shown for the atomic weight distribution in hexagonal boron nitride (h-BN, mp-984). There are 10 first-order graphlet histograms, one for each elemental attribute.
Even at first order, the graphlet histograms retain the full distribution of elemental attributes across all sites in the unit cell. In contrast, composition-only schemes such as MAGPIE~\cite{meredig_combinatorial_2014,ward_general-purpose_2016} compress this information into scalar statistics (mean, min, max, range, mode). However, as Ref.~\cite{lesser_learning_2025} demonstrates, the true power of the graphlet histogram representation enters through capturing the local chemical environment at higher-order graphlets.

\para To capture these higher-order environments, we must first establish a physically rigorous neighbor graph for the crystal.
While GNNs often define connectivity using rigid radial cutoffs or arbitrary $k$-nearest neighbor heuristics, we employ a density-adaptive Voronoi tessellation~\cite{ward_including_2017}.
In this approach, we consider two atoms candidate neighbors if their atomic Wigner--Seitz cells share a common geometric face.
To eliminate spurious, non-physical neighbor connections that can arise from infinitesimally small shared faces in a pure mathematical tessellation, we apply two screening conditions.
First, we retain only neighbors whose Voronoi face weight, defined as the solid angle subtended by a given shared face normalized by the largest such solid angle for that site, exceeds 1\%.
Second, their interatomic distance must be within 1.5 times the sum of their effective atomic radii (Fig.~\ref{fig:graphlets}b), ensuring a physically reasonable scale for orbital overlap.
This yields a robust neighbor map that naturally adapts to the disparate packing fractions found across different materials, serving as the rigid topological foundation for extracting our higher-order graphlets.

\para Second-order graphlets move beyond point properties by defining clusters over all valid connected site pairs from the screened Voronoi graph (illustrated for h-BN in Fig.~\ref{fig:graphlets}c).
For each two-atom cluster, we compute permutation-invariant pair features using the mean and absolute difference of the 10 site attributes, along with the interatomic distance.
Re-emphasizing our departure from black-box message passing, this step yields 21 explicit, physically interpretable second-order distributions that directly capture the diversity of bond-level environments within the unit cell.

\para Third-order graphlets expand the local environment to capture angular constraints.
We define each third-order graphlet as a center site with two valid Voronoi neighbors.
For each such triplet, we compute permutation-invariant summaries---the mean, standard deviation, skewness, and kurtosis---of the 10 site attributes.
Crucially, bond angles first appear at this order and serve as an additional geometric degree of freedom.
Rather than burying this angular geometry in complex line-graph convolutions, we construct 48 distinct, fully transparent third-order distributions.
This yields 79 graphlet distributions per material, capturing local structural geometry in an interpretable, expert-designed representation.

\para
A domain-designed representation is incomplete without its intrinsic distance metric: the rule that determines when two materials are physically ``similar''~\cite{ward_general-purpose_2016,isayev_universal_2017}.
Thoughtful choice of metric drives kernel-based property prediction. More generally, the multiplicity of reported structures under different experimental conditions necessitates systematic methods for resolving these multiplicities.
Obvious choices of pointwise metrics such as Euclidean ($L_2$) distance, Kullback--Leibler divergence, and Jensen--Shannon divergence
respond poorly to small perturbations. Such pointwise metrics
compare bins independently and are blind to the physical proximity of adjacent bins.
A meaningful metric should 
recognize a shift of mass into an adjacent bin as proportionally small, not maximal.
In Ref.~\cite{lesser_learning_2025}, we introduced an Earth Mover's Distance (EMD)-based metric tailored to the graphlet histogram representation; we describe it here in full to make the present paper self-contained.
The EMD, or Wasserstein-1 distance~\cite{rubner_earth_2000} (Fig.~\ref{fig:graphlets}e), measures the minimum work---mass times distance---to transform one distribution into another, penalizing shifts proportionally to their physical magnitude.

\para
For one-dimensional histograms, the EMD has a closed-form solution as the $L_1$ distance between cumulative distributions,
\begin{equation}
    \mathrm{EMD}_i(m_1,m_2) = \Delta_i\sum_{k}\bigl|F(h_1)_{i,k} - F(h_2)_{i,k}\bigr|,
\end{equation}
where $i$ indexes the graphlet feature, $k$ indexes the histogram bin, $\Delta_i$ is the bin width, $F(h)_{i,k}$ is the cumulative sum of the $i$-th histogram through bin $k$, and $m_1$, $m_2$ are any two materials being compared.
The representation assigns each material $N = 10 + 21 + 48 = 79$ distributions; dividing each $\mathrm{EMD}_i$ by its bin width $\Delta_i$ renders the per-feature distances dimensionless and commensurable across features, and their sum defines a scalar inter-material distance,
\begin{equation}
    D(m_1,m_2) = \sum_{i=1}^{N} \frac{\mathrm{EMD}_i(m_1,m_2)}{\Delta_i}.
\end{equation}

\para
Equipped with this metric, we successfully address both motivating challenges.
Because graphlet distributions are normalized, physically equivalent structures---whether reported as a primitive cell or a supercell---yield identical histograms and therefore zero EMD, resolving CIF multiplicity in experimental databases.
Second, as Ref.~\cite{lesser_learning_2025} shows, the per-feature EMDs assemble into a positive-definite additive kernel,
\begin{equation}
    K(m_1,m_2) = \sum_{i=1}^{N} w_i \exp\!\left(-\frac{\mathrm{EMD}_i(m_1,m_2)}{\ell_i}\right),
\end{equation}
where $w_i$ are feature weights and $\ell_i$ are length scales optimized during training.
Positive-definiteness follows from the equivalence between one-dimensional EMD and the $L_1$ distance over cumulative distributions, which makes $\exp(-\mathrm{EMD}/\ell)$ a valid Mercer kernel~\cite{lesser_learning_2025}.
This kernel enabled the GP-$T_c$ model to achieve competitive accuracy for superconducting critical temperature prediction from ${\sim}4{,}350$ experimental structures~\cite{lesser_learning_2025}.

\para
We applied the graphlet featurization described above to the Materials Project~\cite{jain_commentary_2013}, computing histograms up to third order for every material in the database.
Graphlet-MP comprises 149{,}082 materials, each described by 79 distributions (60~GiB total), and is openly available at Ref.~\cite{database}.
These precomputed histograms are ready for immediate use in any downstream learning task, kernel-based or otherwise, without re-featurizing the underlying crystal structures.
The accompanying open-source codebase~\cite{source_code} provides the full pipeline for constructing graphlet histograms from arbitrary CIFs and for computing the EMD metric, enabling researchers to extend the database to new structures or apply the metric framework to their own datasets.

\para
In summary, this paper provides a self-contained specification of the graphlet histogram representation and its associated metric.
From an individual CIF, screened Voronoi tessellation extracts first-order (site), second-order (bond), and third-order (bond-angle) graphlets, yielding 79 interpretable distributions per material.
The Earth Mover's Distance furnishes a perturbation-robust metric over these distributions that resolves CIF multiplicity and underpins kernel-based regression in the low-data regime.
Graphlet-MP, a precomputed database of 149{,}082 materials, and an open-source codebase make both the representation and the metric immediately available for reuse.

\para
We envision Graphlet-MP~\cite{database} serving as a shared foundation: any research group can immediately train property predictors---for superconductivity, dielectrics, piezoelectrics, or other properties governed by local coordination geometry---using the precomputed graphlet features and the EMD metric.
At the same time, the representation extends beyond any single database.
The accompanying open-source codebase~\cite{source_code} applies directly to experimentally determined CIFs from the Inorganic Crystal Structure Database (ICSD), proprietary synthesis campaigns, or targeted high-throughput screens, making it a practical tool for laboratories that generate their own structural data and wish to place new materials in the same representational space.
As the community increasingly recognizes the limitations of data-hungry black-box models for guiding real experimental programs, we anticipate that transparent, precomputable representations like graphlet histograms---backed by both a ready-to-use database and an extensible codebase---will play a central role in bridging computational prediction and laboratory discovery.

{\bf Acknowledgments:} EAK was supported in part by the AI Research Institutes program supported by the NSF and Intel Corporation under NSF award DMR-2433348, AI Materials Institute (AI-MI). The database is hosted on the AI-MI server, supported by the program. AP was supported by the U.S. Department of Energy, Office of Science, Basic Energy Sciences, Materials Sciences and Engineering Division. EAK was supported in part by the U.S. Department of Energy, Office of Science, Basic Energy Sciences, Materials Sciences and Engineering Division.
OL and EAK were supported in part by the U.S. Department of Energy through Award Number: DE-SC0023905. YL and EAK were supported in part by the MURI grant FA9550-21-1-0429.

\bibliography{Graphlets}

\begin{thebibliography}{17}%
\makeatletter
\providecommand \@ifxundefined [1]{%
 \@ifx{#1\undefined}
}%
\providecommand \@ifnum [1]{%
 \ifnum #1\expandafter \@firstoftwo
 \else \expandafter \@secondoftwo
 \fi
}%
\providecommand \@ifx [1]{%
 \ifx #1\expandafter \@firstoftwo
 \else \expandafter \@secondoftwo
 \fi
}%
\providecommand \natexlab [1]{#1}%
\providecommand \enquote  [1]{``#1''}%
\providecommand \bibnamefont  [1]{#1}%
\providecommand \bibfnamefont [1]{#1}%
\providecommand \citenamefont [1]{#1}%
\providecommand \href@noop [0]{\@secondoftwo}%
\providecommand \href [0]{\begingroup \@sanitize@url \@href}%
\providecommand \@href[1]{\@@startlink{#1}\@@href}%
\providecommand \@@href[1]{\endgroup#1\@@endlink}%
\providecommand \@sanitize@url [0]{\catcode `\\12\catcode `\$12\catcode
  `\&12\catcode `\#12\catcode `\^12\catcode `\_12\catcode `\%12\relax}%
\providecommand \@@startlink[1]{}%
\providecommand \@@endlink[0]{}%
\providecommand \url  [0]{\begingroup\@sanitize@url \@url }%
\providecommand \@url [1]{\endgroup\@href {#1}{\urlprefix }}%
\providecommand \urlprefix  [0]{URL }%
\providecommand \Eprint [0]{\href }%
\providecommand \doibase [0]{https://doi.org/}%
\providecommand \selectlanguage [0]{\@gobble}%
\providecommand \bibinfo  [0]{\@secondoftwo}%
\providecommand \bibfield  [0]{\@secondoftwo}%
\providecommand \translation [1]{[#1]}%
\providecommand \BibitemOpen [0]{}%
\providecommand \bibitemStop [0]{}%
\providecommand \bibitemNoStop [0]{.\EOS\space}%
\providecommand \EOS [0]{\spacefactor3000\relax}%
\providecommand \BibitemShut  [1]{\csname bibitem#1\endcsname}%
\let\auto@bib@innerbib\@empty
\bibitem [{\citenamefont {Choudhary}\ and\ \citenamefont
  {DeCost}(2021)}]{choudhary_atomistic_2021}%
  \BibitemOpen
  \bibfield  {author} {\bibinfo {author} {\bibfnamefont {K.}~\bibnamefont
  {Choudhary}}\ and\ \bibinfo {author} {\bibfnamefont {B.}~\bibnamefont
  {DeCost}},\ }\bibfield  {title} {\bibinfo {title} {Atomistic {Line} {Graph}
  {Neural} {Network} for improved materials property predictions},\ }\href
  {https://doi.org/10.1038/s41524-021-00650-1} {\bibfield  {journal} {\bibinfo
  {journal} {npj Computational Materials}\ }\textbf {\bibinfo {volume} {7}},\
  \bibinfo {pages} {185} (\bibinfo {year} {2021})}\BibitemShut {NoStop}%
\bibitem [{\citenamefont {Goodall}\ and\ \citenamefont
  {Lee}(2020)}]{goodall_predicting_2020}%
  \BibitemOpen
  \bibfield  {author} {\bibinfo {author} {\bibfnamefont {R.~E.~A.}\
  \bibnamefont {Goodall}}\ and\ \bibinfo {author} {\bibfnamefont {A.~A.}\
  \bibnamefont {Lee}},\ }\bibfield  {title} {\bibinfo {title} {Predicting
  materials properties without crystal structure: deep representation learning
  from stoichiometry},\ }\href {https://doi.org/10.1038/s41467-020-19964-7}
  {\bibfield  {journal} {\bibinfo  {journal} {Nature Communications}\ }\textbf
  {\bibinfo {volume} {11}},\ \bibinfo {pages} {6280} (\bibinfo {year}
  {2020})}\BibitemShut {NoStop}%
\bibitem [{\citenamefont {Xie}\ and\ \citenamefont
  {Grossman}(2018)}]{xie_crystal_2018}%
  \BibitemOpen
  \bibfield  {author} {\bibinfo {author} {\bibfnamefont {T.}~\bibnamefont
  {Xie}}\ and\ \bibinfo {author} {\bibfnamefont {J.~C.}\ \bibnamefont
  {Grossman}},\ }\bibfield  {title} {\bibinfo {title} {Crystal {Graph}
  {Convolutional} {Neural} {Networks} for an {Accurate} and {Interpretable}
  {Prediction} of {Material} {Properties}},\ }\href
  {https://doi.org/10.1103/PhysRevLett.120.145301} {\bibfield  {journal}
  {\bibinfo  {journal} {Physical Review Letters}\ }\textbf {\bibinfo {volume}
  {120}},\ \bibinfo {pages} {145301} (\bibinfo {year} {2018})}\BibitemShut
  {NoStop}%
\bibitem [{\citenamefont {Sommer}\ \emph {et~al.}(2023)\citenamefont {Sommer},
  \citenamefont {Willa}, \citenamefont {Schmalian},\ and\ \citenamefont
  {Friederich}}]{sommer_3dsc_2023}%
  \BibitemOpen
  \bibfield  {author} {\bibinfo {author} {\bibfnamefont {T.}~\bibnamefont
  {Sommer}}, \bibinfo {author} {\bibfnamefont {R.}~\bibnamefont {Willa}},
  \bibinfo {author} {\bibfnamefont {J.}~\bibnamefont {Schmalian}},\ and\
  \bibinfo {author} {\bibfnamefont {P.}~\bibnamefont {Friederich}},\ }\bibfield
   {title} {\bibinfo {title} {3{DSC} - a dataset of superconductors including
  crystal structures},\ }\href {https://doi.org/10.1038/s41597-023-02721-y}
  {\bibfield  {journal} {\bibinfo  {journal} {Scientific Data}\ }\textbf
  {\bibinfo {volume} {10}},\ \bibinfo {pages} {816} (\bibinfo {year}
  {2023})}\BibitemShut {NoStop}%
\bibitem [{\citenamefont {Lesser}\ \emph {et~al.}(2026)\citenamefont {Lesser},
  \citenamefont {Liu}, \citenamefont {Maus}, \citenamefont {Panigrahi},
  \citenamefont {Mallayya}, \citenamefont {Gong}, \citenamefont {Kabra},
  \citenamefont {Lee}, \citenamefont {Chatterjee}, \citenamefont {Merino},
  \citenamefont {Weinberger}, \citenamefont {Schoop}, \citenamefont {Gardner},\
  and\ \citenamefont {Kim}}]{lesser_learning_2025}%
  \BibitemOpen
  \bibfield  {author} {\bibinfo {author} {\bibfnamefont {O.}~\bibnamefont
  {Lesser}}, \bibinfo {author} {\bibfnamefont {Y.}~\bibnamefont {Liu}},
  \bibinfo {author} {\bibfnamefont {N.}~\bibnamefont {Maus}}, \bibinfo {author}
  {\bibfnamefont {A.}~\bibnamefont {Panigrahi}}, \bibinfo {author}
  {\bibfnamefont {K.}~\bibnamefont {Mallayya}}, \bibinfo {author}
  {\bibfnamefont {A.}~\bibnamefont {Gong}}, \bibinfo {author} {\bibfnamefont
  {A.}~\bibnamefont {Kabra}}, \bibinfo {author} {\bibfnamefont {S.~B.}\
  \bibnamefont {Lee}}, \bibinfo {author} {\bibfnamefont {S.}~\bibnamefont
  {Chatterjee}}, \bibinfo {author} {\bibfnamefont {A.}~\bibnamefont {Merino}},
  \bibinfo {author} {\bibfnamefont {K.~Q.}\ \bibnamefont {Weinberger}},
  \bibinfo {author} {\bibfnamefont {L.~M.}\ \bibnamefont {Schoop}}, \bibinfo
  {author} {\bibfnamefont {J.~R.}\ \bibnamefont {Gardner}},\ and\ \bibinfo
  {author} {\bibfnamefont {E.-A.}\ \bibnamefont {Kim}},\ }\href
  {https://doi.org/10.48550/arXiv.2510.07373} {\bibinfo {title} {Electron
  affinity difference distributions guide the discovery of the superconductor
  {{PtPb}}$_3${{Bi}}}} (\bibinfo {year} {2026}),\ \Eprint
  {https://arxiv.org/abs/2510.07373} {arXiv:2510.07373 [cond-mat.supr-con]}
  \BibitemShut {NoStop}%
\bibitem [{Note1()}]{Note1}%
  \BibitemOpen
  \bibinfo {note} {There are ${\sim }9{,}150$ entries in 3DSC. Out of those,
  6{,}463 are unique superconductors, and only 4{,}325 have unique
  representative structural information, as measured by the graphlet histogram
  earth mover distance metric, as of June 2026~\cite
  {lesser_learning_2025}.}\BibitemShut {Stop}%
\bibitem [{\citenamefont {Wang}\ \emph {et~al.}(2021)\citenamefont {Wang},
  \citenamefont {Kauwe}, \citenamefont {Murdock},\ and\ \citenamefont
  {Sparks}}]{wang_compositionally_2021}%
  \BibitemOpen
  \bibfield  {author} {\bibinfo {author} {\bibfnamefont {A.~Y.-T.}\
  \bibnamefont {Wang}}, \bibinfo {author} {\bibfnamefont {S.~K.}\ \bibnamefont
  {Kauwe}}, \bibinfo {author} {\bibfnamefont {R.~J.}\ \bibnamefont {Murdock}},\
  and\ \bibinfo {author} {\bibfnamefont {T.~D.}\ \bibnamefont {Sparks}},\
  }\bibfield  {title} {\bibinfo {title} {Compositionally restricted
  attention-based network for materials property predictions},\ }\href
  {https://doi.org/10.1038/s41524-021-00545-1} {\bibfield  {journal} {\bibinfo
  {journal} {npj Computational Materials}\ }\textbf {\bibinfo {volume} {7}},\
  \bibinfo {pages} {77} (\bibinfo {year} {2021})}\BibitemShut {NoStop}%
\bibitem [{\citenamefont {Ghiringhelli}\ \emph {et~al.}(2015)\citenamefont
  {Ghiringhelli}, \citenamefont {Vybiral}, \citenamefont {Levchenko},
  \citenamefont {Draxl},\ and\ \citenamefont
  {Scheffler}}]{ghiringhelli_big_2015}%
  \BibitemOpen
  \bibfield  {author} {\bibinfo {author} {\bibfnamefont {L.~M.}\ \bibnamefont
  {Ghiringhelli}}, \bibinfo {author} {\bibfnamefont {J.}~\bibnamefont
  {Vybiral}}, \bibinfo {author} {\bibfnamefont {S.~V.}\ \bibnamefont
  {Levchenko}}, \bibinfo {author} {\bibfnamefont {C.}~\bibnamefont {Draxl}},\
  and\ \bibinfo {author} {\bibfnamefont {M.}~\bibnamefont {Scheffler}},\
  }\bibfield  {title} {\bibinfo {title} {Big {Data} of {Materials} {Science}:
  {Critical} {Role} of the {Descriptor}},\ }\href
  {https://doi.org/10.1103/PhysRevLett.114.105503} {\bibfield  {journal}
  {\bibinfo  {journal} {Physical Review Letters}\ }\textbf {\bibinfo {volume}
  {114}},\ \bibinfo {pages} {105503} (\bibinfo {year} {2015})}\BibitemShut
  {NoStop}%
\bibitem [{\citenamefont {Meredig}\ \emph {et~al.}(2014)\citenamefont
  {Meredig}, \citenamefont {Agrawal}, \citenamefont {Kirklin}, \citenamefont
  {Saal}, \citenamefont {Doak}, \citenamefont {Thompson}, \citenamefont
  {Zhang}, \citenamefont {Choudhary},\ and\ \citenamefont
  {Wolverton}}]{meredig_combinatorial_2014}%
  \BibitemOpen
  \bibfield  {author} {\bibinfo {author} {\bibfnamefont {B.}~\bibnamefont
  {Meredig}}, \bibinfo {author} {\bibfnamefont {A.}~\bibnamefont {Agrawal}},
  \bibinfo {author} {\bibfnamefont {S.}~\bibnamefont {Kirklin}}, \bibinfo
  {author} {\bibfnamefont {J.~E.}\ \bibnamefont {Saal}}, \bibinfo {author}
  {\bibfnamefont {J.~W.}\ \bibnamefont {Doak}}, \bibinfo {author}
  {\bibfnamefont {A.}~\bibnamefont {Thompson}}, \bibinfo {author}
  {\bibfnamefont {K.}~\bibnamefont {Zhang}}, \bibinfo {author} {\bibfnamefont
  {A.}~\bibnamefont {Choudhary}},\ and\ \bibinfo {author} {\bibfnamefont
  {C.}~\bibnamefont {Wolverton}},\ }\bibfield  {title} {\bibinfo {title}
  {Combinatorial screening for new materials in unconstrained composition space
  with machine learning},\ }\href {https://doi.org/10.1103/PhysRevB.89.094104}
  {\bibfield  {journal} {\bibinfo  {journal} {Physical Review B}\ }\textbf
  {\bibinfo {volume} {89}},\ \bibinfo {pages} {094104} (\bibinfo {year}
  {2014})}\BibitemShut {NoStop}%
\bibitem [{\citenamefont {Ward}\ \emph {et~al.}(2016)\citenamefont {Ward},
  \citenamefont {Agrawal}, \citenamefont {Choudhary},\ and\ \citenamefont
  {Wolverton}}]{ward_general-purpose_2016}%
  \BibitemOpen
  \bibfield  {author} {\bibinfo {author} {\bibfnamefont {L.}~\bibnamefont
  {Ward}}, \bibinfo {author} {\bibfnamefont {A.}~\bibnamefont {Agrawal}},
  \bibinfo {author} {\bibfnamefont {A.}~\bibnamefont {Choudhary}},\ and\
  \bibinfo {author} {\bibfnamefont {C.}~\bibnamefont {Wolverton}},\ }\bibfield
  {title} {\bibinfo {title} {A general-purpose machine learning framework for
  predicting properties of inorganic materials},\ }\href
  {https://doi.org/10.1038/npjcompumats.2016.28} {\bibfield  {journal}
  {\bibinfo  {journal} {npj Computational Materials}\ }\textbf {\bibinfo
  {volume} {2}},\ \bibinfo {pages} {16028} (\bibinfo {year}
  {2016})}\BibitemShut {NoStop}%
\bibitem [{\citenamefont {Isayev}\ \emph {et~al.}(2017)\citenamefont {Isayev},
  \citenamefont {Oses}, \citenamefont {Toher}, \citenamefont {Gossett},
  \citenamefont {Curtarolo},\ and\ \citenamefont
  {Tropsha}}]{isayev_universal_2017}%
  \BibitemOpen
  \bibfield  {author} {\bibinfo {author} {\bibfnamefont {O.}~\bibnamefont
  {Isayev}}, \bibinfo {author} {\bibfnamefont {C.}~\bibnamefont {Oses}},
  \bibinfo {author} {\bibfnamefont {C.}~\bibnamefont {Toher}}, \bibinfo
  {author} {\bibfnamefont {E.}~\bibnamefont {Gossett}}, \bibinfo {author}
  {\bibfnamefont {S.}~\bibnamefont {Curtarolo}},\ and\ \bibinfo {author}
  {\bibfnamefont {A.}~\bibnamefont {Tropsha}},\ }\bibfield  {title} {\bibinfo
  {title} {Universal fragment descriptors for predicting properties of
  inorganic crystals},\ }\href {https://doi.org/10.1038/ncomms15679} {\bibfield
   {journal} {\bibinfo  {journal} {Nature Communications}\ }\textbf {\bibinfo
  {volume} {8}},\ \bibinfo {pages} {15679} (\bibinfo {year}
  {2017})}\BibitemShut {NoStop}%
\bibitem [{sou()}]{source_code}%
  \BibitemOpen
  \href@noop {} {}\bibinfo {howpublished}
  {\url{https://github.com/ai-materials-institute/GraphletDatabase}}\BibitemShut
  {NoStop}%
\bibitem [{\citenamefont {Rumble}\ \emph {et~al.}(2020)\citenamefont {Rumble},
  \citenamefont {Bruno},\ and\ \citenamefont {Doa}}]{Rumble2020}%
  \BibitemOpen
  \bibinfo {editor} {\bibfnamefont {J.~R.}\ \bibnamefont {Rumble}}, \bibinfo
  {editor} {\bibfnamefont {T.~J.}\ \bibnamefont {Bruno}},\ and\ \bibinfo
  {editor} {\bibfnamefont {M.~J.}\ \bibnamefont {Doa}},\ eds.,\ \href@noop {}
  {\emph {\bibinfo {title} {{{CRC}} Handbook of Chemistry and Physics: A
  Ready-Reference Book of Chemical and Physical Data}}},\ \bibinfo {edition}
  {101st}\ ed.\ (\bibinfo  {publisher} {CRC Press, Taylor \& Francis Group},\
  \bibinfo {address} {Boca Raton London New York},\ \bibinfo {year}
  {2020})\BibitemShut {NoStop}%
\bibitem [{\citenamefont {Ward}\ \emph {et~al.}(2017)\citenamefont {Ward},
  \citenamefont {Liu}, \citenamefont {Krishna}, \citenamefont {Hegde},
  \citenamefont {Agrawal}, \citenamefont {Choudhary},\ and\ \citenamefont
  {Wolverton}}]{ward_including_2017}%
  \BibitemOpen
  \bibfield  {author} {\bibinfo {author} {\bibfnamefont {L.}~\bibnamefont
  {Ward}}, \bibinfo {author} {\bibfnamefont {R.}~\bibnamefont {Liu}}, \bibinfo
  {author} {\bibfnamefont {A.}~\bibnamefont {Krishna}}, \bibinfo {author}
  {\bibfnamefont {V.~I.}\ \bibnamefont {Hegde}}, \bibinfo {author}
  {\bibfnamefont {A.}~\bibnamefont {Agrawal}}, \bibinfo {author} {\bibfnamefont
  {A.}~\bibnamefont {Choudhary}},\ and\ \bibinfo {author} {\bibfnamefont
  {C.}~\bibnamefont {Wolverton}},\ }\bibfield  {title} {\bibinfo {title}
  {Including crystal structure attributes in machine learning models of
  formation energies via {Voronoi} tessellations},\ }\href
  {https://doi.org/10.1103/PhysRevB.96.024104} {\bibfield  {journal} {\bibinfo
  {journal} {Physical Review B}\ }\textbf {\bibinfo {volume} {96}},\ \bibinfo
  {pages} {024104} (\bibinfo {year} {2017})}\BibitemShut {NoStop}%
\bibitem [{\citenamefont {Rubner}\ \emph {et~al.}(2000)\citenamefont {Rubner},
  \citenamefont {Tomasi},\ and\ \citenamefont {Guibas}}]{rubner_earth_2000}%
  \BibitemOpen
  \bibfield  {author} {\bibinfo {author} {\bibfnamefont {Y.}~\bibnamefont
  {Rubner}}, \bibinfo {author} {\bibfnamefont {C.}~\bibnamefont {Tomasi}},\
  and\ \bibinfo {author} {\bibfnamefont {L.~J.}\ \bibnamefont {Guibas}},\
  }\bibfield  {title} {\bibinfo {title} {The {Earth} {Mover}'s {Distance} as a
  {Metric} for {Image} {Retrieval}},\ }\href
  {https://doi.org/10.1023/A:1026543900054} {\bibfield  {journal} {\bibinfo
  {journal} {International Journal of Computer Vision}\ }\textbf {\bibinfo
  {volume} {40}},\ \bibinfo {pages} {99} (\bibinfo {year} {2000})}\BibitemShut
  {NoStop}%
\bibitem [{\citenamefont {Jain}\ \emph {et~al.}(2013)\citenamefont {Jain},
  \citenamefont {Ong}, \citenamefont {Hautier}, \citenamefont {Chen},
  \citenamefont {Richards}, \citenamefont {Dacek}, \citenamefont {Cholia},
  \citenamefont {Gunter}, \citenamefont {Skinner}, \citenamefont {Ceder},\ and\
  \citenamefont {Persson}}]{jain_commentary_2013}%
  \BibitemOpen
  \bibfield  {author} {\bibinfo {author} {\bibfnamefont {A.}~\bibnamefont
  {Jain}}, \bibinfo {author} {\bibfnamefont {S.~P.}\ \bibnamefont {Ong}},
  \bibinfo {author} {\bibfnamefont {G.}~\bibnamefont {Hautier}}, \bibinfo
  {author} {\bibfnamefont {W.}~\bibnamefont {Chen}}, \bibinfo {author}
  {\bibfnamefont {W.~D.}\ \bibnamefont {Richards}}, \bibinfo {author}
  {\bibfnamefont {S.}~\bibnamefont {Dacek}}, \bibinfo {author} {\bibfnamefont
  {S.}~\bibnamefont {Cholia}}, \bibinfo {author} {\bibfnamefont
  {D.}~\bibnamefont {Gunter}}, \bibinfo {author} {\bibfnamefont
  {D.}~\bibnamefont {Skinner}}, \bibinfo {author} {\bibfnamefont
  {G.}~\bibnamefont {Ceder}},\ and\ \bibinfo {author} {\bibfnamefont {K.~A.}\
  \bibnamefont {Persson}},\ }\bibfield  {title} {\bibinfo {title} {Commentary:
  The {Materials Project}: A materials genome approach to accelerating
  materials innovation},\ }\href {https://doi.org/10.1063/1.4812323} {\bibfield
   {journal} {\bibinfo  {journal} {APL Materials}\ }\textbf {\bibinfo {volume}
  {1}},\ \bibinfo {pages} {011002} (\bibinfo {year} {2013})}\BibitemShut
  {NoStop}%
\bibitem [{dat()}]{database}%
  \BibitemOpen
  \href@noop {} {}\bibinfo {howpublished}
  {\url{https://doi.org/10.5281/zenodo.20532978}}\BibitemShut {NoStop}%
\end{thebibliography}%

\end{document}